\documentclass[twocolumn,prb,aps,floatfix,longbibliography,superscriptaddress]{revtex4-1}
\pdfoutput=1
\usepackage{color}
\usepackage{graphicx}
\graphicspath{{images/}{./}} 
\usepackage{siunitx}
\usepackage[color=green!60,textsize=small]{todonotes}
\usepackage{physics}
\usepackage{amsthm}
\usepackage{amsmath}
\usepackage{amssymb}
\usepackage{enumerate}
\usepackage{placeins}
\usepackage{booktabs}
\usepackage{dsfont}
\usepackage{hyperref}

\AtBeginDocument{%
\heavyrulewidth=.08em
\lightrulewidth=.05em
\cmidrulewidth=.03em
\belowrulesep=.65ex
\belowbottomsep=0pt
\aboverulesep=.4ex
\abovetopsep=0pt
\cmidrulesep=\doublerulesep
\cmidrulekern=.5em
\defaultaddspace=.5em
}

\begin{document}

\title{A Parameter-Free Differential Evolution Algorithm for the Analytic Continuation of Imaginary Time Correlation Functions}

\author{Nathan S. Nichols}
\affiliation{Data Science and Learning Division, Argonne National Laboratory, Argonne, Illinois 60439, USA}
\affiliation{Department of Physics, University of Vermont, Burlington, VT 05405, USA}
\affiliation{Materials Science Program, University of Vermont, Burlington, VT
05404,USA}

\author{Paul Sokol}
\affiliation{Department of Physics, Indiana University, Bloomington, IN 47408, USA}

\author{Adrian Del Maestro}
\affiliation{Department of Physics and Astronomy, University of Tennessee, Knoxville, TN 37996, USA}
\affiliation{Min H. Kao Department of Electrical Engineering and Computer Science, University of Tennessee, Knoxville, TN 37996, USA}

\begin{abstract}
    We report on Differential Evolution for Analytic Continuation (DEAC): a parameter-free evolutionary algorithm to generate the dynamic structure factor from imaginary time correlation functions. Our approach to this long-standing problem in quantum many-body physics achieves enhanced spectral fidelity while using fewer compute (CPU) hours. The need for fine-tuning of algorithmic control parameters is eliminated by embedding them within the genome to be optimized for this evolutionary computation based algorithm.  Benchmarks are presented for models where the dynamic structure factor is known exactly, and experimentally relevant results are included for quantum Monte Carlo simulations of bulk $^4$He below the superfluid transition temperature. 
\end{abstract}

\maketitle


\section{Introduction}
Imaginary time correlation functions can be extended to the real
time domain via analytic continuation \cite{Pathria:2022}. However, the process
to achieve accurate spectral functions from quantum Monte Carlo (QMC) data is a
notoriously ill-posed inverse problem as a result of the stochastic uncertainty
that ensures the equivalent likelihood of many different possible reconstructions through direct inverse Laplace transformations.

Current approaches to the inverse problem are labeled by a zoo of acronyms and
include: the average spectrum method (ASM) \cite{Khaldoon:2020},
stochastic optimization with consistent constraints (SOCC) \cite{Goulko:2017},
genetic inversion via falsification of theories (GIFT) \cite{Bertaina:2017, Vitali:2010},
the famous maximum entropy method (MEM) \cite{Jarrell:1996},
and the fast and efficient stochastic optimization method (FESOM) \cite{Bao:2016}.
The ASM approach performs a functional average over all admissible spectral functions,
while SOCC uses random updates to the spectrum consistent with error bars on the input data.
The GIFT method uses a genetic algorithm with many algorithmic control parameters.
The traditionally used MEM utilizes Bayesian inference and is further described
in Section~\ref{ssec:MEM}. Finally, a state-of-the-art approach FESOM adds
random noise to proposed spectra at each iteration, averaging the spectra when a
level of fitness is reached, and is further described in Section~\ref{ssec:FESOM}.
More recent work has focused on applying machine learning techniques \cite{Fournier:2020, Raghavan:2021, Kades:2020} with limited success for particular types of imaginary time correlations.

In this paper we introduce a new method: the differential evolution for analytic
continuation (DEAC) algorithm, to achieve reconstructed dynamic structure
factors $S(\vb*{q},\omega)$ from the imaginary time intermediate scattering
function. Similar to the GIFT method \cite{Bertaina:2017}, a population of
candidate spectral functions is maintained whose average fitness is improved
through recombination over several generations. Control parameters are adjusted
using self adaptive techniques. This new method is validated against nine
multi-peak spectra at finite temperature and compared with two other robust
and commonly used methods, MEM and FESOM. Our algorithm performs well in terms of speed, accuracy, and ease of use. These three strengths unlock new avenues for scientific discovery through greater utilization of computational resources and better
fidelity of the reconstructed spectra.

The remainder of this paper is organized as follows.  We begin with a comprehensive description of the inverse problem, the construction of simulated quantum Monte Carlo data, and the method used to generate imaginary time intermediate scattering data for superfluid $^4$He. Details are then given for our implementation of the two competing approaches, before proceeding to a detailed discussion of our evolutionary algorithm.  A careful comparison of the results and performance of the DEAC algorithm with both the maximum entropy and stochastic optimization methods is provided for simulated data sets containing ubiquitous spectral features. Moving beyond simulated data, results are shown for the bulk $^4$He spectrum below the superfluid transition temperature. We conclude with an analysis of the resulting spectral functions and discussion of the advantages of each method. Scripts and data used in analysis and plotting as well as details to download the source code for the three analytic continuation methods explored are available online \cite{paper-code-repo}.


\section{Model and Data}
\label{sec:model_and_data}

\subsection{The Inverse Problem}
\label{ssec:inverse_problem}

The dynamic structure factor is a measure of particle correlations in space and
time \cite{Pathria:2022}. It is defined as the temporal Fourier transform:
\begin{equation}
    S(\vb*{q},\omega)=\frac{1}{2\pi}\int_{-\infty}^{\infty}F(\vb*{q},t)
        \exp(i \omega t) \mathrm{d}t
\label{eq:dynamic_structure_factor}
\end{equation}
for wave vector $\vb*{q}$ and frequency $\omega$, where
$F(\vb*{q},t)=\frac{1}{N}\langle\rho_{\vb*{q}}(t)\rho_{\bar{\vb*{q}}}(0)
\rangle$ is the intermediate scattering function, which can be written more
explicitly as
\begin{equation}
    F(\vb*{q},t) = \frac{1}{N}\expval{ \sum_{j,l} e^{-i\vb*{q}\cdot\vb*{r}_j(t)} e^{i\vb*{q}\cdot\vb*{r}_l(0)}}
\end{equation}
in units with Planck's constant $\hbar = 1$ and the Boltzmann constant $k_B = 1$
for time-dependent particle positions $\vb*{r}(t)$.

A determination of the intermediate scattering function in imaginary time is
found by using the detailed balance condition of the dynamic structure factor
$S(\vb*{q}, \omega)=S(\vb*{q}, -\omega)e^{\beta\omega}$, a Wick rotation
of $F(t)$ to $F(-i\tau)$, and a Fourier transform of Eq.~(\ref{eq:dynamic_structure_factor})
giving
\begin{equation}
    F(\vb*{q},\tau) = \int_0^\infty S(\vb*{q},\omega) \qty[e^{-\tau\omega} + e^{-(\beta - \tau)\omega}] d\omega
\label{eq:intermediate_scattering_function}
\end{equation}
for imaginary time $\tau$ and $\beta=\frac{1}{T}$.
Exact results within statistical uncertainties for $F(\vb*{q},\tau)$ can be
produced via quantum Monte Carlo (QMC) simulations \cite{Allen:2017}.

Accurate reconstruction of $S(\vb*{q},\omega)$ through an inverse Laplace
transform of Eq.~(\ref{eq:intermediate_scattering_function}) is problematic. A
brute force approach quickly reveals the ill-conditioned nature of the
transformation and unique solutions are not guaranteed due to the finite
uncertainty in the measured $F(\vb*{q},\tau)$. Furthermore, the use of
periodic boundary conditions in simulations to reduce finite size effects
further restricts measurements to specific momenta
\begin{equation}
\vb*{q} = \sum_{\alpha=1}^{D}\frac{2\pi n_\alpha}{L_\alpha}\,  \hat{e}_\alpha,
\label{eq:qvecs}
\end{equation}
which are commensurate with the periodicity of the $D$-dimensional hypercubic
system with volume $\prod_{\alpha=1}^{D} L_{\alpha}$, $n_{\alpha} \in \mathbb{Z}$
and $\hat{e}_{\alpha}$ denote unit vectors.
The use of incommensurate $\vb*{q}$ vectors results in large deviations from
expected results, especially at low momenta \cite{Villamaina:2014}.  In order to
make comparison with experimental measurements that depend only on the magnitude
of the momentum vector, $q$ (such as with neutron scattering experiments on
powder or liquid samples), we separate results for $S(\vb*{q},\omega)$ into bins
of $[q, q + \Delta q]$ where $\Delta q$ is an arbitrarily chosen spectral
resolution.  Some finite error is introduced with this approach due to the
nonuniform distribution of the magnitudes of $q$ vectors in each bin, but is
mitigated with increasing box size approaching the thermodynamic limit.
Approaches to generating accurate $S(\vb*{q},\omega)$ are discussed in
Section~\ref{sec:methods}.

Spectral moments of integration \cite{Sturm:1993hv}
\begin{equation}
    \langle\omega^k\rangle = \int_{0}^\infty \omega^k S(\vb*{q},\omega)\qty[1+ (-1)^k e^{-\beta\omega}] \mathrm{d}\omega
\label{eq:spectral_moments}
\end{equation}
can be used to reduce the search over the number of possible spectral functions
in some cases.
The inverse first frequency moment $\langle\omega^{-1}\rangle$ is proportional
to the static linear density response function \cite{Pines:1970} and is fixed by
$F(\vb*{q},\tau)$
\begin{equation}
    \langle \omega^{-1} \rangle = \frac{1}{2}\int_0^\beta \mathrm{d}\tau F(\vb*{q},\tau)\, , 
\label{eq:inverse_first_moment}
\end{equation}
while the zeroth frequency moment
\begin{equation}
    \langle\omega^0\rangle = S(\vb*{q})
    \equiv \int_{0}^\infty S(\vb*{q},\omega)(1 + e^{-\beta\omega}) \mathrm{d}\omega
\label{eq:zeroeth_moment}
\end{equation}
is the static structure factor $S(\vb*{q})$ by definition.

These moments of integration are useful when they are exactly known, such as in
the case of neutral quantum liquids where the first frequency moment is
equivalent to the free particle dispersion \cite{Pines:1970, Miller:1962}
\begin{equation}
    \langle\omega^1\rangle = \frac{\hbar \lvert \vb*{q} \rvert ^2}{2m}
\label{eq:first_moment}
\end{equation}
shown here in dimensionful units, or when they can be accurately estimated,
such as in the case of the uniform electron gas or other hard-core gasses for the
third frequency moment \cite{Groth:2019, Puff:1965, Mihara:1968, Iwamoto:1986, Iwamoto:1984}.
To further highlight the general utility and \emph{knowledge free} nature of the
evolutionary algorithm discussed here, we chose to not enforce the moments of
integration. However, their inclusion, when available, could serve to further
enhance the accuracy of the DEAC method. 

While we are ultimately interested in the dynamic structure factor,
$S(\vb*{q},\omega)$, it can be useful to perform the analytic continuation on a
modified kernel by replacing $S(\vb*{q},\omega)$ with $S'(\vb*{q},\omega)$ in
Eqs.~(\ref{eq:intermediate_scattering_function}) and (\ref{eq:spectral_moments})
and transforming back after performing the analytic continuation.
Three useful kernels of integration were determined and are described below. 
The standard kernel,
$S'(\vb*{q},\omega)=S(\vb*{q},\omega)$,
is simply the dynamic structure factor.
The normalization kernel,
$S'(\vb*{q},\omega)=S(\vb*{q},\omega)(1 + e^{-\beta\omega})$,
simplifies the static structure factor \cite{Boninsegni:1996}.
The hyperbolic kernel
$S'(\vb*{q},\omega)=2S(\vb*{q},\omega)e^{-\frac{\beta\omega}{2}}$,
severely constrains the modified intermediate scattering function while causing
hyperbolic terms to appear in 
Eqs.~(\ref{eq:intermediate_scattering_function}) and (\ref{eq:spectral_moments}).
These kernels exhibit different performance in terms of CPU hours, but give
generally similar resulting spectra. The hyperbolic kernel was used with the
simulated QMC data and the normalization kernel was used to produce the bulk
$^4$He spectrum.


\subsection{Simulated Quantum Monte Carlo Data}
\label{ssec:simulated_data}
\begin{table*}[t]
    \renewcommand{\arraystretch}{1.5}
    \setlength\tabcolsep{12pt}
    \begin{tabular}{@{}lllllll@{}} 
        \toprule
        Alias & $p_l$ & $\mu_l\; \qty[\si{\kelvin}]$ & $\sigma_l\; \qty[\si{\kelvin}]$ & $p_r$ & $\mu_r\; \qty[\si{\kelvin}]$ & $\sigma_r\; \qty[\si{\kelvin}]$\\
        \midrule
        same height close (shc)       & 0.50 & 15.0 & 3.0 & 0.50 & 35.0 & 3.0 \\
        same height far (shf)         & 0.50 & 15.0 & 3.0 & 0.50 & 45.0 & 3.0 \\
        same height overlapping (sho) & 0.50 & 15.0 & 3.0 & 0.50 & 25.0 & 3.0 \\
        short tall close (stc)        & 0.25 & 15.0 & 3.0 & 0.75 & 35.0 & 3.0 \\
        short tall far (stf)          & 0.25 & 15.0 & 3.0 & 0.75 & 45.0 & 3.0 \\
        short tall overlapping (sto)  & 0.25 & 15.0 & 3.0 & 0.75 & 25.0 & 3.0 \\
        tall short close (tsc)        & 0.75 & 15.0 & 3.0 & 0.25 & 35.0 & 3.0 \\
        tall short far (tsf)          & 0.75 & 15.0 & 3.0 & 0.25 & 45.0 & 3.0 \\
        tall short overlapping (tso)  & 0.75 & 15.0 & 3.0 & 0.25 & 25.0 & 3.0 \\
        \bottomrule
    \end{tabular}
    \caption{\label{tab:dsf_parameters} Parameters to generate the dynamic structure factor 
    and intermediate scattering function from Eqs.~(\ref{eq:dsf_exact}) and
    (\ref{eq:isf_exact}) respectively. The subscripts $l$
    and $r$ correspond to the left-most peak and right-most peak of the spectral
    function in the positive frequency space.}
\end{table*}
Simulated quantum Monte Carlo data was generated to determine how each of the
methods described in the next section perform at reconstructing
$S(\vb*{q},\omega)$ from $F(\vb*{q},\tau)$. A data set of spectral functions
\begin{equation}
    S_\mathrm{exact}(\vb*{q},\omega)= p_l\tilde{s}(\omega,\mu_l,\sigma_l,\beta) + p_r\tilde{s}(\omega,\mu_r,\sigma_r,\beta)
    \label{eq:dsf_exact}
\end{equation}
was created from a superposition of two Gaussian-like features of the form
\begin{equation}
    \tilde{s}(\omega,\mu,\sigma,\beta)=\Theta(\omega)s(\omega,\mu,\sigma) +
    \Theta(-\omega)s(-\omega,\mu,\sigma)e^{-\beta\omega}
    \label{eq:dsf_gaussian_like}
\end{equation}
scaled by a factor $p_{l|r}$ where 
\begin{equation}
    s(x,\mu,\sigma)=\frac{1}{\sigma\sqrt{2\pi}}e^{-\frac{1}{2}\bigl(\frac{x - \mu}{\sigma}\bigr)^2}
\label{eq:gaussian}
\end{equation}
is a normalized Gaussian function centered at $\mu$ with width determined by
$\sigma$. The spectra were normalized by their respective static structure
factors, $S(\vb*{q})$.
The exact intermediate scattering function for such spectra can be calculated
using Eq.~(\ref{eq:intermediate_scattering_function}) as
\begin{equation}
    \tilde{F}_\mathrm{sim}(\vb*{q},\tau) = p_l\tilde{f}(\tau,\mu_l,\sigma_l,\beta) + p_r\tilde{f}(\tau,\mu_r,\sigma_r,\beta)
    \label{eq:isf_exact}
\end{equation}
where
\begin{equation}
    \tilde{f}(\tau,\mu,\sigma,\beta) = \frac{1}{2} e^{-\frac{\mu^2}{2 \sigma^2}}\qty[
    f(\tau,\mu,\sigma) + f(\beta - \tau,\mu,\sigma)]
    \label{eq:isf_gaussian_like}
\end{equation}
and
\begin{equation}
    f(x,\mu,\sigma) = e^{\frac{(\mu - x\sigma^2)^2}{2 \sigma^2}} \qty{
    1 + 
\text{erf}\qty[\frac{1}{\sigma\sqrt{2}} (\mu - x\sigma^2)]}\, .
\end{equation}
The first frequency moment for a single Gaussian-like spectra
$\tilde{s}(\omega,\mu,\sigma,\beta)$ can be calculated via Eq.~(\ref{eq:spectral_moments})
as
\begin{multline}
    \expval{\omega^1_{\tilde{s}}} = \frac{1}{2}\left \{e^{\frac{\beta^2 \sigma^2}{2} - \beta \mu} \qty(\beta \sigma^2 - \mu) \text{erfc}\qty[\frac{ \qty(\beta
   \sigma^2 - \mu)}{\sigma\sqrt{2}}] \right. \\
   + \left. \mu \qty[\text{erf}\qty(\frac{\mu}{\sigma\sqrt{2}})+1] \right\}.
    \label{eq:first_moment_exact}
\end{multline}

Simulated quantum Monte Carlo data $F_\mathrm{sim}(\vb{q},\tau)$ was generated by adding normally
distributed noise to the exact intermediate scattering function for $N_s = 1000$
samples and averaging the results: 
\begin{equation}
    F_\mathrm{sim}(\vb*{q},\tau) = \frac{1}{N_s}\sum_1^{N_s} (1 + \epsilon \mathcal{N}(0,1)) \tilde{F}_\mathrm{sim}(\vb*{q},\tau)
    \label{eq:simulated_data}
\end{equation}
where $\mathcal{N}$ is the standard normal distribution and $\epsilon$ is the
noise amplitude. Three separate noise amplitudes were explored
$\epsilon={0.0001,0.001,0.01}$ and are referred to as small, medium, and
large error. These labels are not intended as commentary on the quality of the
simulated data and are only used for easy reference between the three error
levels.

Nine simulated data sets at each error level were generated with two
Gaussian-like peaks in the positive frequency space. Parameters to
Eqs.~(\ref{eq:dsf_exact}) and (\ref{eq:isf_exact}) used to generate the exact
dynamic structure factors and intermediate scattering functions along with
aliases for each spectra are found in Table~\ref{tab:dsf_parameters}. These
parameters were chosen to simulate experimentally relevant spectra and explore
the resolving power of spectral features for each analytic continuation method
explored. Each data set was generated at temperature $T=1.2\,\si{\kelvin}$ for
$M=129$ imaginary time steps from $\tau_0 = 0$ to $\tau_M = \frac{\beta}{2}$.


\subsection{Bulk Helium Quantum Monte Carlo Data}
\label{ssec:qmc_data}
Liquid helium is the most accessible and best studied strongly interacting quantum fluid \cite{Glyde:2017, Feynman:1955, Pines:2018a, Pines:2018b}.  A demonstration of the ability of the DEAC algorithm to generate experimentally relevant spectra will be presented in Section~\ref{ssec:qmc_data_results} by reproducing the phonon-roton spectrum of bulk $^4$He from quantum Monte Carlo data. The results presented herein utilize our open source path integral quantum Monte Carlo code in the canonical ensemble (access details in Ref.~\onlinecite{delmaestroSVN}). Simulations were performed with temperature $T=1.35\,\si{\kelvin}$, chemical potential $\mu=-5.47\,\si{\kelvin}$, $M=100$ imaginary time slices, and $N=64$ particles.  The finite box size was determined by setting the density corresponding to saturated vapor pressure with $L_x=L_y=L_z\approx 14.31158\,\si{\angstrom}$ \cite{Herdman:2014zu}.  For the helium-helium interactions we adopted the Aziz intermolecular potential \cite{Aziz:1979}.  Data was collected for 1357 different $\vb*{q}$ vectors constructed according to Eq.~(\ref{eq:qvecs}) corresponding to all allowable vectors with magnitudes $q\le 3.0\,\si[per-mode = power]{\per\angstrom}$.  Imaginary time symmetry around $\tau=\frac{\beta}{2}$ was used to combine measurements taken for $F(\vb*{q},{\beta}/{2} + i\Delta\tau)$ and $F(\vb*{q},{\beta}/{2} - i\Delta\tau)$.  Results for each vector were jackknife averaged over 100 separate seeds.


\section{Previous Methods}
\label{sec:methods}

\subsection{Maximum Entropy Method}
\label{ssec:MEM}
The standard and most commonly used approach for determining spectral functions from imaginary time
correlation functions is the Maximum Entropy Method (MEM) \cite{Jarrell:1996,
Bergeron:2016}.  Bayesian inference is used to optimize the likelihood function
and prior probability. Starting from Bayes' theorem:
\begin{equation}
    P(S|F) = \frac{P(F|S)P(S)}{P(F)}
\end{equation}
where $P(S|F)$ is the probability of obtaining spectrum $S$ given the intermediate scattering function
function $F$, $P(F|S)$ is the probability of obtaining $F$ given $S$ and is the
\textit{likelihood}, $P(S)$ is the so called \textit{prior probability} of
obtaining spectrum $S$, and $P(F)$ is a marginal probability that can be ignored
in this treatment as it is constant. 
Through the central limit theorem, a proportionality can be determined for the
likelihood
\begin{equation}
    P(F|S)\propto e^{-\frac{1}{2}\chi^2}
\end{equation}
where
\begin{equation}
    \chi^2 = \sum_{i=0}^M \frac{1}{M}\frac{(F_i - \bar{F}_i)^2}{\sigma_i^2}.
    \label{eq:chi_squared}
\end{equation}
The expected value $\bar{F}_i$ is the averaged simulation data, while $\sigma_i^2$ is
its variance at imaginary time slice $\tau_i$, and $M$ is the number of imaginary time slices.

A form for the prior probability that obeys the properties of the spectral
function can be introduced to constrain the search space for possible solutions
\begin{equation}
    P(S) \propto e^{\alpha \hat{S}}
\end{equation}
where $\alpha$ is the regularization constant and the information gain (or 
relative entropy term)
\begin{equation}
    \hat{S} = -\sum_i \frac{\Delta\omega_i}{2\pi}S(\omega_i) \ln \frac{S(\omega_i)}{D(\omega_i)}.
    \label{eq:entropy_like_term}
\end{equation}
is the Kullback-Leibler divergence of a spectral function $A(\omega)$
from some default model $D(\omega)$ that captures prior information of the
spectrum after discretization of the frequency space.

The posterior probability can then be described by
\begin{equation}
    P(S|F)\propto e^{\alpha \hat{S} -\frac{1}{2}\chi^2}.
\end{equation}
Maximizing this quantity amounts to the minimization of 
\begin{equation}
    Q[S] = \frac{1}{2}\chi^2 - \alpha \hat{S}.
    \label{eq:MEM_fitness}
\end{equation}
We use the Broyden–Fletcher–Goldfarb–Shanno (BFGS) method as the algorithm
of choice \cite{Malouf:2002, Galen:2007} to minimize Eq.~(\ref{eq:MEM_fitness}).
A maximum of 20000 BFGS iterations are performed for each simulation.

There are several approaches to determining the appropriate regularization
constant $\alpha$ and we employ a recent method developed by Bergeron and Tremblay
\cite{Bergeron:2016}.
\begin{figure}[t]
    \centering
    \includegraphics[width=\columnwidth]{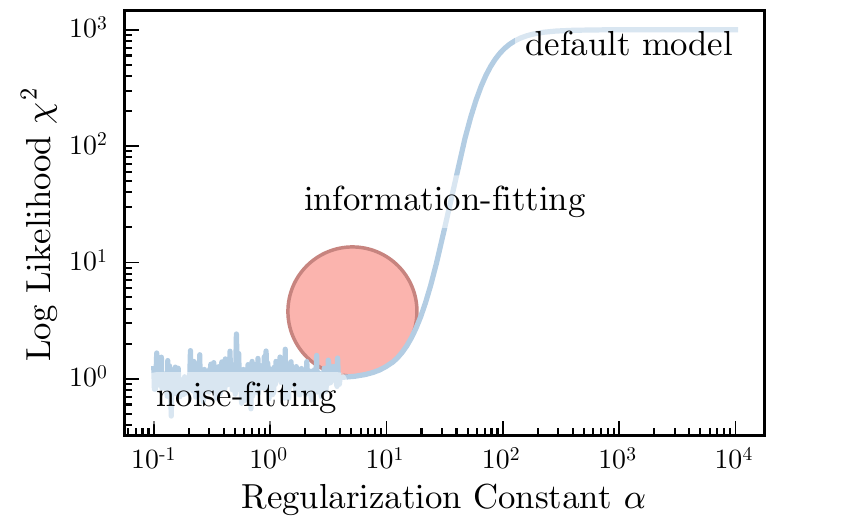}
    \caption{\label{fig:maxent_cartoon} A schematic representation of the method
    to determine the optimal regularization constant when using the maximum
    entropy approach. The noise-fitting region is characterized by little to no
    change in the recovered spectra with changes to $\alpha$ and is dominated by
    fitting to the noise in the intermediate scattering function. The
    information-fitting region corresponds to spectra with deviations from the
    default model strongly affected by $\alpha$. The default model region consists
    of spectra with little or no deviation from the default model. An example of a
    circle fit to the data is shown in red, where an estimate of curvature can
    be made made from the radius.
    }
\end{figure}
A schematic representation of their approach is shown in Figure~\ref{fig:maxent_cartoon}
where a sweep over possible $\alpha$ is performed with the optimal value
chosen by computing the curvature of $\log_{10} \chi^2$ as a function of
$\log_{10} \alpha$. Curvature is estimated as $\kappa=\frac{1}{R}$ where $R$ is
the radius of a circle fit to the data. Three distinct regions are observed:
noise-fitting, information-fitting, and default model region. The $\alpha$
value corresponding to the maximum curvature close to the noise-fitting region
is the value that recovers the optimal spectral function.

The default model $D(\omega)$ was chosen to be a single Gaussian-like peak in
the positive frequency space using Eq.~(\ref{eq:dsf_gaussian_like}) for an
equally spaced frequency partition of size $N=129$ ranging from
$\omega_0=0.0\,\si{\kelvin}$ to
$\omega_N=64.0\,\si{\kelvin}$. For each model per simulated spectra,
the parameter $\mu$ was chosen to be the first moment as calculated by Eq.~(\ref{eq:first_moment_exact})
and the parameter: 
\begin{equation}
    \sigma = \frac{\mathrm{min}\qty(\expval{\omega^1}- \omega_0,\, \omega_N - \expval{\omega^1})}{3}.
    \label{eq:sigma}
\end{equation}
where the denominator allows for sufficient damping of the default model
before reaching the edges of the frequency search space.
The initial guess for $S(\vb*{q},\omega)$ for each MEM simulation was set to the
default model. The regularization constant $\alpha$ was swept over for an equally spaced
partition in $\log_{10}$-space of size $N_\alpha=1001$ from $10^{-1}$ to $10^4$.
Optimal final spectra at each error level for the data set described by
Table~\ref{tab:dsf_parameters} were determined as described above. 


\subsection{Fast and Efficient Stochastic Optimization Method}
\label{ssec:FESOM}
The fast and efficient stochastic optimization method (FESOM) \cite{Bao:2016} is a state-of-the-art
technique to determine spectral functions from imaginary time quantum Monte Carlo data. The
approach uses minimal prior information by solely optimizing the likelihood
function. This is achieved by brute force minimization of Eq.~(\ref{eq:chi_squared}) through a
numerical algorithm (described below) to within an acceptable tolerance level
$\eta$. Several FESOM simulations are performed and the final spectrum is
determined by averaging the results. A confidence band can be constructed by
taking the standard deviation. This treatment of the final spectrum is
statistically allowable since each realization has the same posterior
probability when $\chi^2 = \eta$.

In practice, a FESOM simulation is performed as follows. An initial spectrum is
generated on a discretized frequency space
$\{\omega_0 \le \omega_1 \le \ldots \le \omega_i \le \ldots \le \omega_{N-1} \le \omega_N \}$
obeying the normalization condition Eq.~(\ref{eq:zeroeth_moment}). The quality
of fit $\chi^2$ is calculated via Eq.~(\ref{eq:chi_squared}). For each iteration,
an update to the spectrum is proposed by scaling each spectral weight
$S(\vb*{q},\omega_i)$ by $\lvert1 + x_i\rvert$ where $x_i\in\mathcal{N}(0,1)$ and
normalizing by the static structure factor $S(\vb*{q})$.  If the new spectrum
has a $\chi^2$ value that is smaller than the previous iteration, the update is
accepted. Iterations are performed until acceptable tolerance is achieved as
described previously. In our simulations, the initial spectrum was generated
using the same method described for the default model above in
Section\ref{ssec:MEM} with the exception that the frequency space partition
size was $N=513$. For each error level, $N_R = 1000$ reconstructions of the
spectral function were measured using FESOM to an acceptable tolerance level of
$\eta=5\epsilon$. A maximum of $N_I=10^7$ iterations were performed for each
simulation. The results were averaged to generate a final spectrum at each error
level for the data set described by Table~\ref{tab:dsf_parameters}. The final
spectra were smoothed by averaging the spectral weight in adjacent frequency bins.

\section{Differential Evolution for Analytic Continuation}
\label{sec:DEAC}

\begin{figure*}[t]
    \centering
    \includegraphics[width=\textwidth]{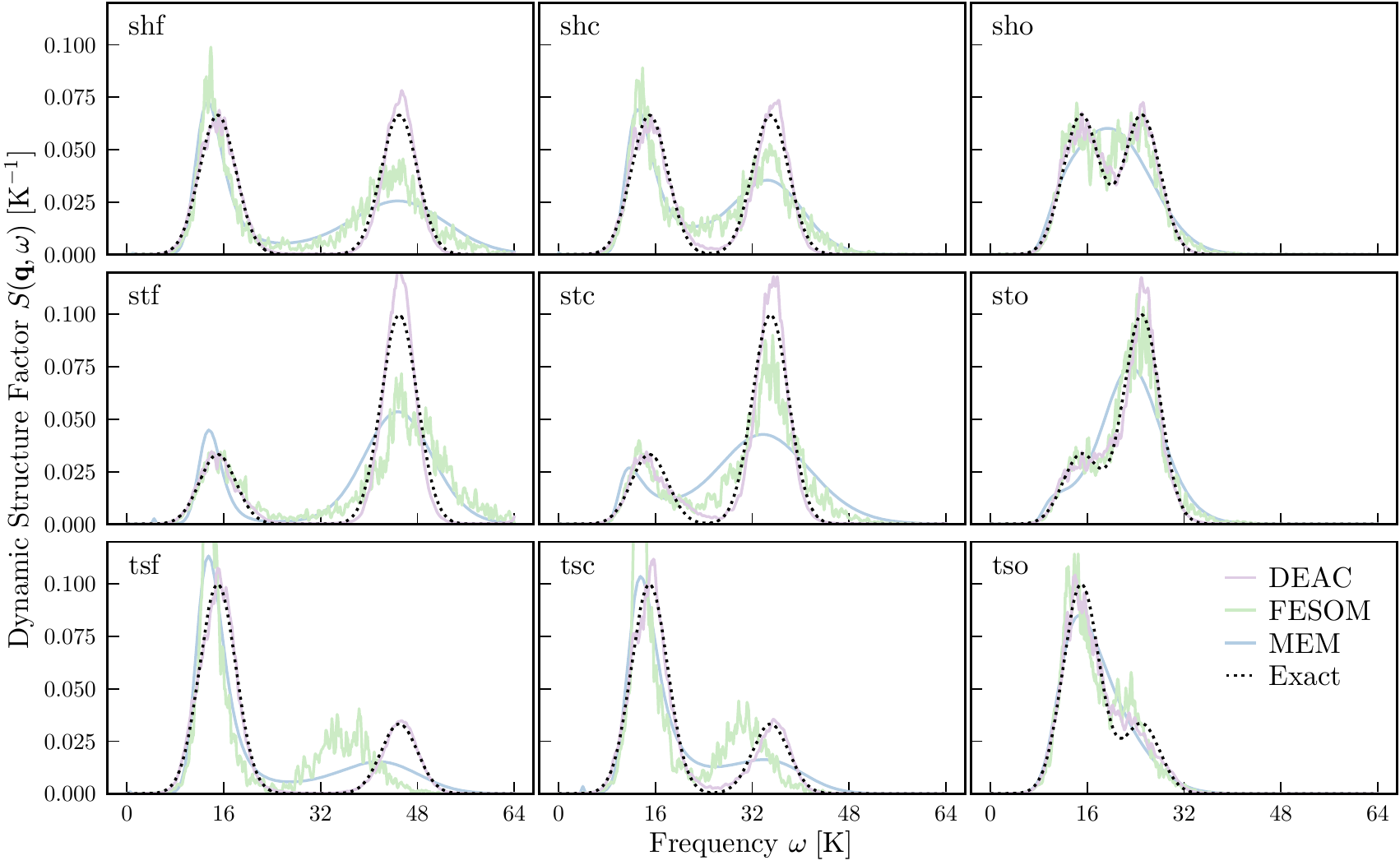}
    \caption{\label{fig:nine_panel_small} Nine different spectral
    reconstructions using different analytic continuation methods on simulated
    quantum Monte Carlo data. The black dashed line is the exact spectra. Results
    are shown for the smallest error level explored $\epsilon=0.0001$. The DEAC
    algorithm accurately captures the spectral features in all cases explored.
    }
\end{figure*}
Inspired by the genetic inversion via falsification of theories (GIFT)
algorithm \cite{Bertaina:2017, Vitali:2010}, we developed an approach using
evolutionary computation, the differential evolution for analytic continuation (DEAC)
method that does not rely on hyperparameters. A comparison of GIFT with MEM at
$T=0\;\si{\kelvin}$ can be found in Ref.~\cite{Kora:2018}.
The new approach developed here expands to the more difficult
finite temperature regime and uses an evolutionary computation method well
suited for a genome consisting of real valued numbers.

Differential evolution \cite{StornPrice:1997}
is a class of evolutionary algorithms, which determines an optimal solution
within a certain tolerance based on fitness criteria. A population of candidate
solutions is maintained and updated through a simple vector process described
below. As the simulation progresses, each candidate solution is rated and added
to the population based on some fitness criteria and the average fitness of the
population improves. Here, the population is comprised of spectral functions
$S(\vb*{q},\omega)$ discretized over a fixed frequency space, and the fitness of
candidate solutions was calculated via Eq.~(\ref{eq:chi_squared}).

Each iteration of the DEAC algorithm generates a new candidate population by the
following process. For each agent $\tilde{S}_m(\vb*{q},\omega)$ in the
population, three other agents $\tilde{S}_j$, $\tilde{S}_k$, and $\tilde{S}_l$
are randomly chosen such that $\tilde{S}_m \ne \tilde{S}_j \ne \tilde{S}_k \ne \tilde{S}_l$.
A potential new member $\tilde{S}_n$ is created by iterating over the frequency
space $\{\omega_0,\ldots,\omega_N\}$ and for each $\omega_i$ 
\begin{equation}
    \tilde{S}_n(\omega_i)=
    \begin{cases}
        \tilde{S}_j(\omega_i) + \gamma [\tilde{S}_k(\omega_i) - \tilde{S}_l(\omega_i)],& \mathcal{U}(0,1) \leq P^c\\
        \tilde{S}_m(\omega_i),              & \text{otherwise}
    \end{cases}
    \label{eq:DEAC_mutation}
\end{equation}
where $\mathcal{U}(0,1)$ is a random number drawn from the standard uniform
distribution, $P^c$ is the crossover probability, and $\gamma$ is the differential
weight. The new agent $\tilde{S}_n$ replaces $\tilde{S}_m$ in the next generation
if fitness improves over $\tilde{S}_m$, otherwise $\tilde{S}_m$ is retained.

In a standard differential evolution simulation, the differential weight and the
crossover probability would need to be optimized, and while in principle they
should not affect the final outcome, in practice, a poor choice can affect
convergence. Here we employ a self adaptive approach \cite{Qin:2005, Fan:2016}
by embedding $\gamma$ and $P_c$ within the genome of the candidate solutions,
such that each $\tilde{S}_m$ has a corresponding $P^c_m$ and $\gamma_m$. Updates
to the crossover probability are performed 10\% of the time by
\begin{equation}
    P^c_n=
    \begin{cases}
        \mathcal{U}(0,1),& \mathcal{U}(0,1) \leq 0.1\\
        P^c_m,& \text{otherwise}
    \end{cases}
\end{equation}
and updates to the differential weight are also performed 10\% of the time by
\begin{equation}
    \gamma_n=
    \begin{cases}
        \mathcal{U}(0,2),&  \mathcal{U}(0,1) \leq 0.1\\
        \gamma_m,& \text{otherwise}
    \end{cases}\, .
\end{equation}
Note that the control parameters are updated before generating a new candidate
population and $P^c_n$ and $\gamma_n$ should be used in Eq.~(\ref{eq:DEAC_mutation}).
This ensures that beneficial changes to the crossover probability or 
differential weight are preserved.

The population size $N_P$ can be as small as $N_P=4$ and as large as the
computing resource can manage. An optimal solution can be reached for any
$N_P \geq 4$, although scaling of $N_P$ can help determine a population size
that conforms to system constraints and an acceptable usage of CPU hours. Here
we use $N_P = 16$ for the simulated data, and $N_P=8$ for the bulk helium data.

A maximum of $N_I=10^7$ iterations were performed, where the average fitness of
the candidate solution population improved with each generation. Once an
individual solution $\tilde{S}_m(\vb*{q},\omega)$ reaches an acceptable
tolerance level $\chi^2=\eta$, the simulation is terminated and the solution
returned as the optimal spectra $S(\vb*{q},\omega)$. The tolerance levels were
chosen to be the same as those used in FESOM (Section~\ref{ssec:FESOM}) for the
simulated data, and $\eta=1.0$ for superfluid helium.  The frequency space
partition size was $N=513$ and ranged from $\omega_0=0.0\,\si{\kelvin}$ to
$\omega_N=64.0\,\si{\kelvin}$ for the simulated data, and $N=4096$ ranging
from $\omega_0=0.0\,\si{\kelvin}$ to $\omega_N=512.0\,\si{\kelvin}$ for helium.
In each case, $N_R = 1000$ reconstructions of the spectral function were
measured. The results were averaged to generate a final spectrum at each error
level for the data set described by Table~\ref{tab:dsf_parameters} and for each
wave vector examined for the helium. The final spectra were smoothed by
averaging the spectral weight in adjacent frequency bins. Similar to FESOM,
confidence bands can be generated by taking the standard deviation.


\section{Results}
\label{sec:results}

\subsection{Benchmarking on Simulated Data}
\label{ssec:simulated_data_results}
Reconstructed spectra found using DEAC, MEM, and FESOM on simulated quantum
Monte Carlo data are shown in Figure~\ref{fig:nine_panel_small} for the \emph{small}
error level, $\epsilon=0.0001$.
The DEAC algorithm achieves improved spectral feature resolution over the other
two methods in all cases.
These improvements can be seen in the goodness of fit calculated as the
lack-of-fit sum of squares
\begin{equation}
    \label{eq:lof}
    \varphi_\mathrm{lof} = \frac{1}{N} \sum_i^N (S(\vb*{q},\omega_i) - S_\mathrm{exact}(\vb*{q},\omega_i))^2
\end{equation}
where squared deviations of the spectral weight at each frequency are averaged.
Across the range of sample data, DEAC achieves the best score (where lower is better)
for all nine benchmarks except in the \emph{tso} case for \emph{medium} error.
DEAC shows almost an order of magnitude of improvement in the goodness of fit
over the other two methods at \emph{small} error as shown in Table~\ref{tab:gof}
and half that at other error levels.
%
\begin{table}[b]
    \renewcommand{\arraystretch}{1.3}
    \setlength\tabcolsep{8pt}
    \begin{tabular}{@{}lc@{  }c@{  }cc@{  }c@{  }cc@{  }c@{  }c@{  }}
        \toprule
        & \multicolumn{3}{c}{small} & \multicolumn{3}{c}{medium} & \multicolumn{3}{c}{large} \\
        Alias & D & F & M & D & F & M & D & F & M \\
        \midrule
        shf & 5.30 & 3.95 & 3.75 & 4.66 & 3.93 & 4.19 & 4.17 & 3.91 & 3.57 \\
        shc & 5.09 & 3.99 & 3.89 & 4.47 & 4.00 & 3.76 & 4.23 & 3.97 & 3.78 \\
        sho & 4.90 & 4.46 & 4.12 & 4.58 & 4.49 & 4.19 & 4.51 & 4.44 & 3.69 \\
        stf & 4.86 & 3.66 & 3.74 & 5.17 & 3.65 & 3.23 & 4.94 & 3.64 & 3.19 \\
        stc & 4.73 & 4.03 & 3.51 & 4.98 & 4.03 & 3.48 & 4.79 & 3.94 & 3.46 \\
        sto & 4.67 & 4.53 & 4.02 & 4.57 & 4.48 & 4.23 & 4.83 & 4.51 & 4.03 \\
        tsf & 4.89 & 3.44 & 3.92 & 4.76 & 3.45 & 4.04 & 3.93 & 3.44 & 3.27 \\
        tsc & 5.12 & 3.63 & 4.09 & 4.31 & 3.63 & 4.13 & 4.06 & 3.62 & 3.48 \\
        tso & 4.69 & 4.23 & 4.47 & 4.21 & 4.20 & 4.50 & 4.48 & 4.12 & 4.22 \\
        \bottomrule
    \end{tabular}
    \caption{\label{tab:gof}
        Logarithmically scaled goodness of fit for each spectral function
        reconstruction at each error level. Values shown are
        $-\log_{10}(\varphi_\mathrm{lof})$ where
        $\varphi_\mathrm{lof}$ is calculated by Eq.~\eqref{eq:lof}. More
        positive values indicate better qualities of fit. Here we abbreviate the
        methods DEAC, FESOM, and MEM as D, F, and M respectively.
    }
\end{table}

\begin{figure}[h]
    \centering
    \includegraphics[width=\columnwidth]{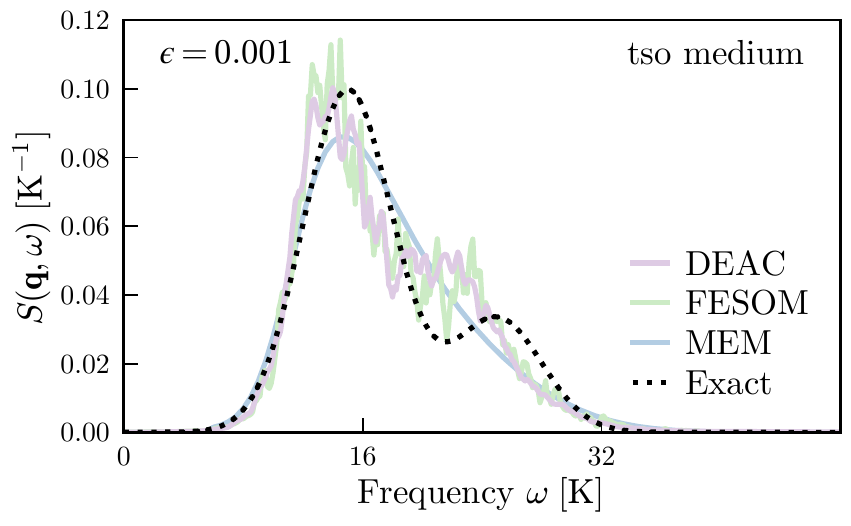}
    \caption{\label{fig:tso_medium}
    Analytic continuation results using DEAC, FESOM, and MEM for the \emph{tso}
    case at \emph{medium} error level. In this case, MEM achieves a better
    goodness of fit over FESOM and DEAC.
    The striking lack or even hint of two spectral peaks for the
    MEM results is reason to judge the FESOM and DEAC results as qualitatively
    better.
    }
\end{figure}

A closer look at the outlier as shown in Figure~\ref{fig:tso_medium} reveals that
although MEM has a better goodness of fit, it lacks the ability to resolve two
distinct peaks. Both FESOM and DEAC indicate a shoulder of a smaller peak next
to the main spectral feature and encourage further QMC data collection to reduce
the error level and achieve better spectral resolution. The perhaps more surprising
result is MEM not winning across all the \emph{close} cases as the method we
employed included prior knowledge by including the first moment as a part of the
default spectrum.

\begin{figure}[t]
    \centering
    \includegraphics[width=\columnwidth]{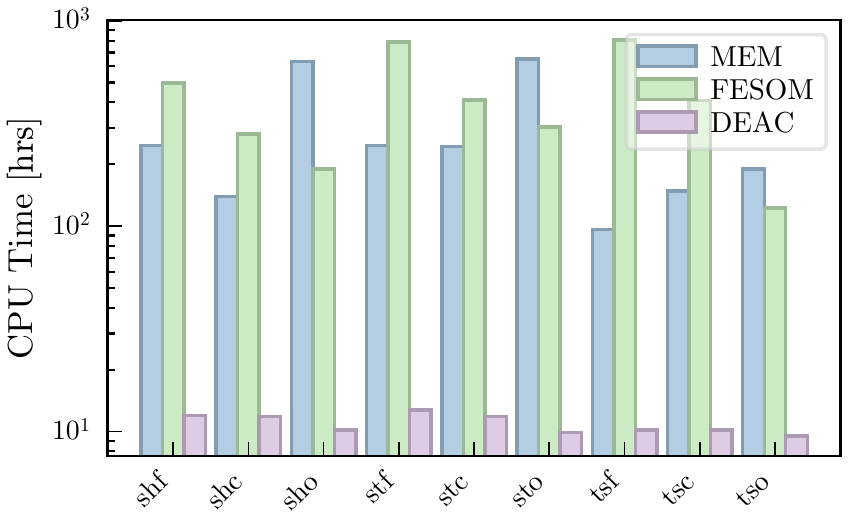}
    \caption{\label{fig:CPU_time_large} CPU time required to generate the final
    spectra using each analytic continuation technique. Timings displayed are
    for the largest error level explored $\epsilon=0.01$. Lower is better, and significant improvement in
    CPU time can be achieved using DEAC. Almost two orders of magnitude
    improvement can be seen over the other methods.
    }
\end{figure}

Another important factor is the computational efficiency of algorithms, as often
spectra must be generated for a large number of $\vb*{q}$ values. The total CPU
time to achieve the final spectra for the 
\emph{large} error data set $\epsilon=0.01$ is shown in Figure~\ref{fig:CPU_time_large}. These
timings include the full parameter sweep for the MEM method and the $N_R=1000$
reconstructions for the FESOM and DEAC methods. They do not include the time
needed to generate the final spectra using the curvature technique for MEM or averaging
the spectra for DEAC and FESOM (as these contributions were negligible).
The MEM results used up to $66\times$ ($14\times$ on average across all
benchmarks) more CPU hours than DEAC, and the FESOM results used up to
$79\times$ ($13\times$ on average across all benchmarks) more CPU hours than DEAC.

\subsection{Bulk Helium}
\label{ssec:qmc_data_results}
\begin{figure}[t]
    \centering
    \includegraphics[width=1\columnwidth]{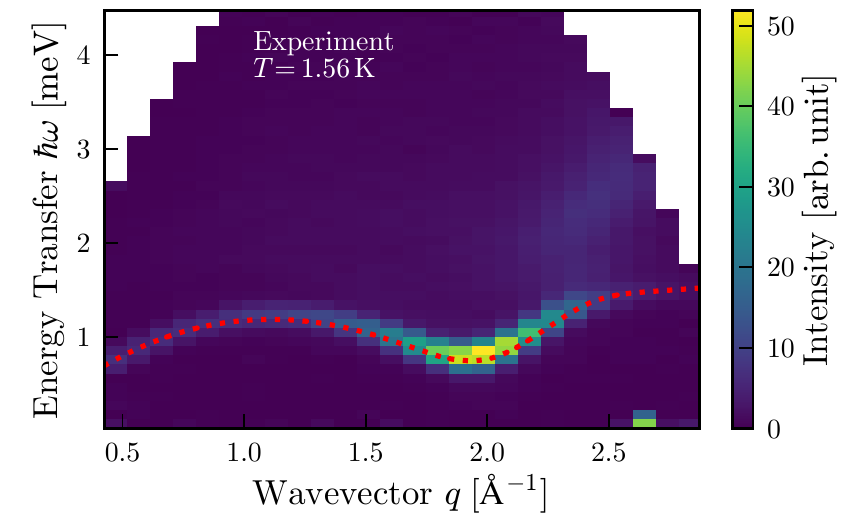}
    \includegraphics[width=1\columnwidth]{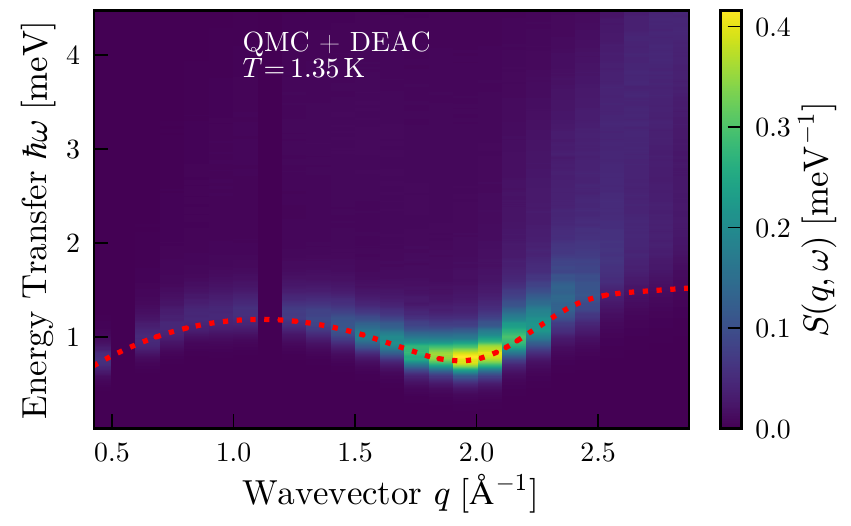}
    \caption{\label{fig:bulk_he_spectra} 
    The phonon-roton spectrum of $^4$He at
    $T=1.56\,\si{\kelvin}$ from neutron scattering experiments on superfluid helium (top) \cite{Prisk:2021} and at
    $T=1.35\,\si{\kelvin}$ as generated by DEAC from canonical quantum Monte Carlo data (bottom).
    Peak locations for experimental measurements of helium at temperatures below
    the superfluid transition temperature $T_\lambda$ and saturated vapor
    pressure are shown as the red dashed line using splines from Donnelly and
    Barenghi \cite{Donnelly:1998}. Good agreement is observed for the maxon and 
    roton locations. Deviations from the experimental spectra and gaps in data
    are due to finite size effects.
    }
\end{figure}

To test the performance of our parameter free algorithm in an experimentally
relevant setting, we consider the well known phonon-roton spectrum of $^4$He
at $T=1.35\,\si{\kelvin}$.  The imaginary time scattering function was generated
from canonical quantum Monte Carlo as described in Section~\ref{ssec:qmc_data}. 
The resulting spectrum in Figure~\ref{fig:bulk_he_spectra} is consistent with
experimental results \cite{Glyde:2017, Donnelly:1998, Godfrin:2021, Godfrin:2018, Prisk:2021},
and we note it involves no
adjustable parameters. Spectral peaks in the maxon and roton regions are found
at momenta $q\approx 1.1\,\si[per-mode = power]{\per\angstrom}$ and
$q\approx2.0\,\si[per-mode = power]{\per\angstrom}$ with energy transfers of
$\omega\approx 1.2\,\si{\milli\electronvolt}$ and
$\omega\approx 0.8\,\si{\milli\electronvolt}$ respectively.
Parts of the linear dispersing branch are observable, but obscured due to
vertical gaps in the spectral data from certain momenta not being measured.
This was either from being incommensurate with periodic boundary conditions or
finite size effects.
Many attempts have been performed to resolve this spectra where much of the
focus has been on a few fixed $q$-values \cite{Boninsegni:1996, Kora:2018, Vitali:2010, Ferre:2016}.
 
\begin{figure}[t]
    \centering
    \includegraphics[width=\columnwidth]{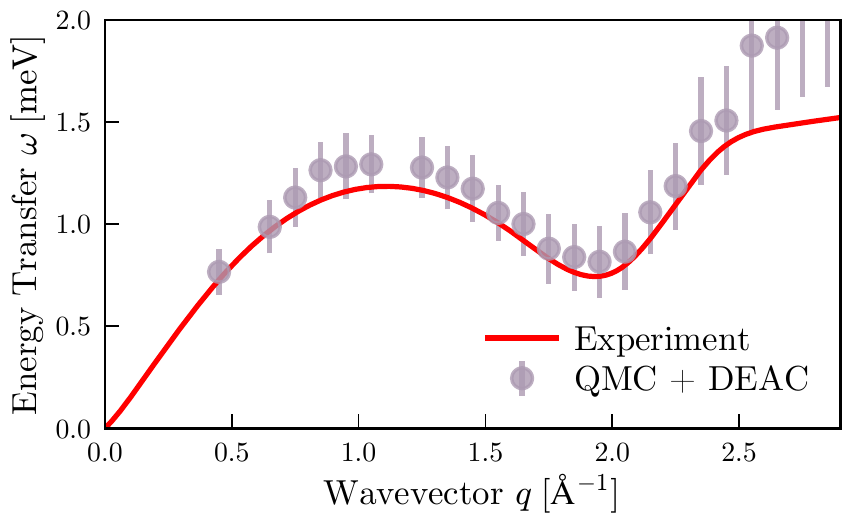}
    \caption{\label{fig:helium_dispersion}
    Maximum peak locations for the $^4$He dispersion at $T=1.35\,\si{\kelvin}$
    as generated by analytic continuation of QMC data using the DEAC algorithm.
    Experimental data for low temperature helium $T < 2.17\,\si{\kelvin}$ at saturated vapor pressure
    is shown \cite{Donnelly:1998}. Again, deviations from the experimental data are
    observed due to finite size effects.
    }
\end{figure}

An advantage that DEAC and FESOM have over other methods is the ability to
estimate confidence bands on spectral features.
For each $N_R$ reconstruction
of the spectrum, we determined the location of the maxima in frequency space and
binned the results for each wavevector $\vb*{q}$ investigated. Then the average
and standard error were calculated in the usual way from the binned data.
In Figure~\ref{fig:helium_dispersion},
we show the average maximum peak locations of the helium dispersion as
determined using standard techniques from the average data including standard
error, where the error bars indicate the full width half maximum.
%


\section{Discussion}
The maximum entropy (MEM) approach is well supported in the literature and performs well for resolving spectral features for well separated peak locations. However, for
closely spaced peaks, it tends to average out the resulting spectra.
This effect can be seen in the \emph{overlapping} cases in Figure~\ref{fig:nine_panel_small}.  Also, for the other non-overlapping cases, there appears to be some skew in the first peak and large broadening of the second peak.

The fast and efficient stochastic optimization approach (FESOM) \cite{Bao:2016}
was able to resolve spectral features in all cases, but had
difficulty in determining the second peak location for the \emph{tall short}
benchmark.  This method was prone to becoming stuck in local optima and not reaching the
selected tolerance level before the maximum number of iterations. For
this reason and for a fairer comparison between the three methods, the timing
results shown are for the \emph{large} error cases where all runs were able to
achieve convergence within the set tolerance. Broadening of the second peak was also an issue for this method.

The differential evolution for analytic continuation (DEAC) algorithm
introduced here provided the best results in the shortest amount of CPU time
in all cases tested. Proof of principle for the ability of this new method to
produce experimentally relevant spectra is shown by the bulk $^4$He spectrum in
Figure~\ref{fig:bulk_he_spectra}. The observed finite size effects can be mitigated
by larger simulation size.

The benchmark spectra were generated using versions of DEAC, FESOM, and MEM
that utilize multithreading, and were written in Julia with source code available
online \cite{deac-jl, fesom-jl, mem-jl}.
Additionally, the bulk helium spectrum was generated by a C\texttt{++} version of DEAC
with optional GPU acceleration (both HIP and CUDA supported) with source code also
available online \cite{deac-cpp}. The authors recommend the C\texttt{++} version of DEAC
over the Julia version.

A note of caution is offered for using any of the three methods described above.
We noticed while exploring the analytic continuation problem that spectral
weight will build up in the final frequency bin if a large enough maximum
frequency is not explored. For the benchmark data, this resulted in the second
peak being pushed to lower energies. This problem is solved by increasing the
maximum frequency at the expense of either CPU time or frequency resolution $\Delta\omega$.

In conclusion, a fast, accurate, and parameter-free method to reconstruct the
dynamic structure factor from imaginary time pair correlation functions has been
developed. The differential evolution for analytic continuation algorithm
uses evolutionary computation with a self adaptive approach to tackle this
long standing problem in many-body physics. Benchmarks on finite temperature
simulated quantum Monte Carlo data against the traditional maximum entropy
method  and the state-of-the-art fast and efficient stochastic optimization
method have shown several advantages. These are found in massive
speedups and the increased fidelity of resulting spectra. The greater ability to
resolve spectral features coupled with reduced computational overhead offers
further opportunity to compare the stochastically exact results from quantum
Monte Carlo with experimental data obtained on the real frequency axis.


\section{Acknowledgments}

We benefited from discussion with H. Barghathi.  This research was supported in
part by the National Science Foundation (NSF) under award Nos.~DMR-1809027 and
DMR-1808440.  Computations were performed on the Vermont Advanced Computing Core
supported in part by NSF award No.~OAC-1827314. This work used the Extreme
Science and Engineering Discovery Environment (XSEDE) \cite{xsede}, which is
supported by NSF grant number ACI-1548562. XSEDE resources used include Bridges and Bridges-2
at Pittsburgh Supercomputing Center, Comet at San Diego Supercomputer Center, and Open
Science Grid (OSG) \cite{osg1, osg2} through allocations TG-DMR190045 and
TG-DMR190101.  OSG is supported by the NSF under award No.~1148698,
and the U.S. Department of Energy's Office of Science.


\FloatBarrier

\nocite{apsrev41Control}
\bibliographystyle{apsrev4-1}
\bibliography{refs}

\end{document}